\documentstyle[12pt]{article}

\begin{document}


\hoffset = -1truecm
\voffset = -2truecm

\title{\bf
Centri`fugal' Force around a Black Hole
}

\author{
{\bf
A.Y.Shiekh
}\\
\normalsize International Centre for Theoretical Physics, Trieste 34100,
Italy
}

\date{6th November 1992}

\maketitle

\begin{abstract}
Besides having some very interesting perturbatively unstable
orbits, it seems that for a Schwarzschild black hole, below $r=3M$,
the force always increases inward with increasing angular
momentum. Here this previously known result is derived with greater
simplicity, and a similar analysis is performed for black holes with
angular momentum and charge.
\end{abstract}



\section{Introduction}

  Much of the time invested in learning physics is dedicated to
developing an intuition for the particular topic in hand. This may
involve the introduction of such imaginary concepts as force,
which may later need to be abandoned (as in quantum theory and
Einstein's theory of general relativity).

  Relativity theory, being so far from the Newtonian physics of the
everyday world, is filled by what initially seems paradoxical
behaviour. General relativity is especially difficult in this respect,
and the partial restoration of gravitational forces (as opposed to
geodesic motion) can help simplify understanding. However,
preconceived notions of force can cause confusion, and it might be
better to deal directly with the concept of the rate of change of
momentum. Indeed, as we shall see, this notion of force is {\it not}, in
general, minus the potential gradient.

  It was this line of reasoning that led Abramowicz and others [1-3]
to investigate the concept of force around a black hole. In the  work that
follows, a particularly direct derivation of some of those  results is
presented, and as a consequence of the simplicity, considerable generalization
was found to be possible.

  It was discovered that, in general, around a black hole there is a
radius below which the force increases inward, ever more so with
increasing angular momentum. This is in conflict with normal
Newtonian intuition, where a spinning object has a natural
tendency to fly off. This interpretation augments an
understanding of the accretion mechanism of black holes and so
may help in identifying certain astronomical observations as being
the result of black hole physics.

  Since the physics of this effect is present even for the
Schwarzschild black hole, the approach is first illustrated for this
case before being applied more generally to a black hole with
charge and angular momentum.

\subsection{Schwarzschild Black Hole}

  Starting directly from the equation of motion, derived from
$p_\mu p^\mu +m^2=0$:

\begin{equation}
\left( {{{dr} \over {d\tau }}}
\right)^2=\mathord{\buildrel{\lower3pt\hbox{$\scriptscriptstyle
\frown$}}\over E} ^2-
\mathord{\buildrel{\lower3pt\hbox{$\scriptscriptstyle\frown$}}
\over V} ^2\left( {\tilde L,\hat r} \right)
\end{equation}

one may derive the radial force:

\begin{equation}
F^r\equiv {{dp^r} \over {d\tau }}
\end{equation}

where:

$$p^r\equiv m{{dr} \over {d\tau }}$$

  This yields the perhaps unexpected relation:

\begin{equation}
{M \over m}F^r=-{1 \over 2}{\partial  \over {\partial \hat
r}}\mathord{\buildrel{\lower3pt\hbox{$\scriptscriptstyle\frown$
}}\over V} ^2
\end{equation}

which reduces to the usual gradient formula far from the hole,
when $\mathord{\buildrel{\lower3pt\hbox{$\scriptscriptstyle\frown$}}
\over V} \sim 1$, where:

\begin{equation}
\mathord{\buildrel{\lower3pt\hbox{$\scriptscriptstyle\frown$
}}\over V} =\sqrt {\left( {1-{2 \over {\hat r}}} \right)\left(
{1+{{\tilde L^2} \over {\hat r^2}}} \right)}
\end{equation}

\centerline{ \it \bf \small
Schwarzschild effective potential (dead black hole, $S = Q = 0$)}

$$
\begin{array}{lllll}
&\mathord{\buildrel{\lower3pt\hbox{$\scriptscriptstyle\frown$
}}\over E} \equiv {E \over m}
&\mathord{\buildrel{\lower3pt\hbox{$\scriptscriptstyle\frown$
}}\over V} \equiv {V \over m}
&\hat r\equiv {r \over M}
&\tilde L\equiv {L \over {Mm}}
\end{array}
$$

as given in Misner, Thorne and Wheeler, 1973 [4], where:

\vskip 0.5cm

{\small
        $E$     is the total (conserved) energy

        $L$     is the (conserved) angular momentum of the test
particle around the black hole

        $M$     is the mass of the black hole

        $V$     is the effective potential

        $m$     is the rest mass of the test particle

        $r$     is the distance of the test particle from the black hole centre

        $F^r$   is the radial force

        $p$     is the momentum of the test particle
}

\vskip 0.5cm

  The effective potential for the Schwarzschild black hole plots out
as:

\setlength{\unitlength}{0.240900pt}
\ifx\plotpoint\undefined\newsavebox{\plotpoint}\fi
\sbox{\plotpoint}{\rule[-0.175pt]{0.350pt}{0.350pt}}%


\centerline{ \it \bf \small
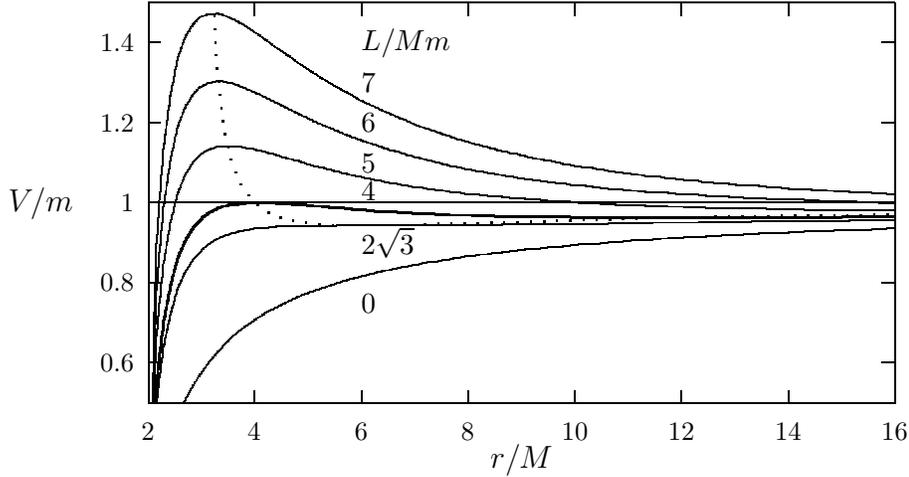
Plot of the effective potential against radius for various
angular momenta
}

\vskip 0.5cm

  The perturbatively unstable orbit at $r=4M$ is especially
interesting, since it could, in theory, be visited and stabilized
(through feedback) without a large expenditure of energy, and so
would be that appropriate for a space-station.

  Then there is the storage aspect. One would assemble a ring of
matter at the $r=4M$ balance, and then, as one sped it up, one
would move it into the corresponding balance between $3M$ and
$4M$. To extract the energy later, one would simply perturb part of
the ring outward, from where it would then fly out to infinity on
its own accord. (There is an energy storage limit of $Mc^2/2$ when
the event horizon rises to the ring orbit).

  Such an accumulator would be well suited if one wished to take
up the energy from a spinning black hole, which might then be
used to kick a space-craft across space. Of course, another such
black hole would be needed to stop or turn about! Gravitational
acceleration is also the only known way to achieve large speed
changes in small times without experiencing destructive forces.

  It is an amazing design of nature. Such orbits, being
perturbatively unstable, would be clear of debris and thus clear
for use by people -- {\it the ultimate sling-shot!}

  One may now derive the central force from:

$${M \over m}F^r=-{1 \over 2}{\partial  \over {\partial \hat
r}}\mathord{\buildrel{\lower3pt\hbox{$\scriptscriptstyle\frown$
}}\over V} ^2$$

to yield:

\begin{equation}
{M \over m}F^r=-\overbrace {{1 \over {\hat r^2}}}^{Static}-\
\overbrace {{{3-\hat r} \over {\hat r^4}}\tilde
L^2}^{Centri`fugal'}
\end{equation}

\centerline{\it \bf \small Schwarzschild force}

\vskip 0.5cm

  For orbits above $3M$, the radial force always finally gets to point
outward, for high enough angular momentum, which is like the
usual Newtonian behaviour. For orbits below $r=3M$ the
`Abramowicz force' [1, 2] always increases inward with rotation.
Here we also see the original discovery [1] that at $r=3M$, the
force is independent of the motion.

  So the phenomena of reversed centri`fugal' force is seen below
$3M$. The results generalise quite simply for a general black hole,
when care is taken to derive the force correctly.

  When investigating accretion, it is perhaps better to contemplate
the situation of conserved angular momentum, rather than fixed
radius (which would require rocket intervention below $r=3M$).

\subsection{General, Kerr-Newman Black Hole (Charged and Spinning):}

  Begin from the equation of equatorial motion [4], derived from
$p_\mu p^\mu+m^2=0$:

\begin{equation}
\left( {p^r} \right)^2={{\alpha E^2-2\beta E+\gamma } \over
{r^4}}
\end{equation}

where:

$$p^r=m{{dr} \over {d\tau }}$$

$$\alpha \equiv \left( {r^2+a^2} \right)^2-a^2 \Delta $$

$$\beta \equiv \left( {La+eQr} \right)\left( {r^2+a^2} \right)-
La\Delta$$

$$\gamma \equiv \left( {La+eQr} \right)^2-L^2\Delta -
m^2r^2\Delta $$

$$\Delta \equiv r^2-2Mr+a^2+Q^2$$

  As before, the force is the rate of change of momentum, and as
such is not always minus the potential gradient. In this case the
force has no direct relation to the effective potential, but is given
by:

\begin{equation}
F^r={1 \over {2m}}{\partial  \over {\partial r}}\left( {{{\alpha
E^2-2\beta E+\gamma } \over {r^4}}} \right)
\end{equation}

and yields the radial force components:

\begin{equation}
\begin{array}{lll}
{M \over m}F^r\left( {\hat a,\hat q} \right)= \\
 -\
\overbrace {{{1-
\mathord{\buildrel{\lower3pt\hbox{$\scriptscriptstyle\frown$}}
\over e}
\mathord{\buildrel{\lower3pt\hbox{$\scriptscriptstyle\frown$}}
\over E} \hat q} \over {\hat r^2}}+{{\left( {1-
\mathord{\buildrel{\lower3pt\hbox{$\scriptscriptstyle\frown$}}
\over E} ^2} \right)\hat a^2+\left( {1-
\mathord{\buildrel{\lower3pt\hbox{$\scriptscriptstyle\frown$}}
\over e} ^2} \right)\hat q^2} \over {\hat
r^3}}+3{{\mathord{\buildrel{\lower3pt\hbox{$\scriptscriptstyle
\frown$}}\over e}
\mathord{\buildrel{\lower3pt\hbox{$\scriptscriptstyle\frown$}}
\over E} \hat q-
\mathord{\buildrel{\lower3pt\hbox{$\scriptscriptstyle\frown$}}
\over E} ^2} \over {\hat r^4}}\hat
a^2+2{{\mathord{\buildrel{\lower3pt\hbox{$\scriptscriptstyle
\frown$}}\over E} ^2} \over {\hat r^5}}\hat a^2\hat q^2}^{Static} \\
  \
-\overbrace {\left( {-{1 \over {\hat r^3}}+{3 \over {\hat r^4}}-
{{2\hat q^2} \over {\hat r^5}}} \right)\tilde L^2}^{Centri`fugal'}-
\overbrace {\left(
{3{{\mathord{\buildrel{\lower3pt\hbox{$\scriptscriptstyle
\frown$}}\over e} \hat q-
2\mathord{\buildrel{\lower3pt\hbox{$\scriptscriptstyle\frown$}}
\over E} } \over {\hat
r^4}}+{{4\mathord{\buildrel{\lower3pt\hbox{$\scriptscriptstyle
\frown$}}\over E} \hat q^2} \over {\hat r^5}}} \right)\hat a
\tilde L}^{Coriolis}
\end{array}
\end{equation}

\centerline{\it \bf \small Equatorial radial force}

\vskip 0.5cm

  The centri`fugal' component changes sign at the radius:

$${{3+\sqrt {9-8\hat q^2}} \over 2}$$

the Coriolis{\footnote{This is not the usual Coriolis force,
but an analogous concept.}} force at:

$${4 \over
3}{{\mathord{\buildrel{\lower3pt\hbox{$\scriptscriptstyle\frown
$}}\over E} \hat q^2} \over
{2\mathord{\buildrel{\lower3pt\hbox{$\scriptscriptstyle\frown$}
}\over E} -
\mathord{\buildrel{\lower3pt\hbox{$\scriptscriptstyle\frown$}}
\over e} \hat q}}$$

  However it is perhaps better to locate the zero point of the
complete dynamical component of the force, that is Coriolis +
centri`fugal'. This stands at the radius:

\begin{equation}
{3 \over 2}\left( {1-\left(
{2\mathord{\buildrel{\lower3pt\hbox{$\scriptscriptstyle\frown$}
}\over E} -
\mathord{\buildrel{\lower3pt\hbox{$\scriptscriptstyle\frown$}}
\over e} \hat q}
\right){{\hat a} \over {\tilde L}}} \right)+\sqrt {{9  \over 4}\left( {1-\left(
{2\mathord{\buildrel{\lower3pt\hbox{$\scriptscriptstyle\frown$}
}\over E} -
\mathord{\buildrel{\lower3pt\hbox{$\scriptscriptstyle\frown$}}
\over e} \hat q}
\right){{\hat a} \over {\tilde L}}} \right)^2-2\left(  {1-
2\mathord{\buildrel{\lower3pt\hbox{$\scriptscriptstyle\frown$}}
\over E} {{\hat a} \over {\tilde L}}} \right)\hat q^2}
\end{equation}

the position of the event horizon being given by:

$$\hat r_{event\  horizon}=1+\sqrt {1-\hat a^2-\hat q^2}$$

where:

$$
\begin{array}{lll}
&\mathord{\buildrel{\lower3pt\hbox{$\scriptscriptstyle\frown$
}}\over E} \equiv {E \over m}
&\mathord{\buildrel{\lower3pt\hbox{$\scriptscriptstyle\frown$
}}\over e} \equiv {e \over m}
\end{array}
$$

$$
\begin{array}{llll}
&\hat r\equiv {r \over M}
&\hat q\equiv {Q \over M}
&\hat a\equiv {a \over M}
\end{array}
$$

$$\tilde L\equiv {L \over {Mm}}$$

as given in Misner, Thorne and Wheeler [4], where:

\vskip 0.5cm

{ \small
        $E$     is the total (conserved) energy

        $L$     is the (conserved) angular momentum of the test
particle around the black hole

        $M$     is the mass of the black hole

        $Q$     is the electric charge of the black hole

        $S$     is the angular momentum of the black hole ($a \equiv S/M$)

        $e$     is the electric charge of the test particle

        $m$     is the rest mass of the test particle

        $r$     is the distance of the test particle from the black hole
centre

        $F^r$   is the radial force

        $p$     is the momentum of the test particle
}

\vskip 0.5cm

\subsection{Appendix}
  Much of the simplicity so far depended upon the symmetries
present. Time symmetry yielded constant energy ($E$), and $\phi$
symmetry yielded conserved angular momentum ($L$). However a
further (Carter) symmetry has been located, and this is equivalent
to having solved the final equation of motion (cf. the
Hamilton-Jacobi approach). The now decoupled equations of motion
are given by:

$$
\begin{array}{llllll}
&p^r={{\sqrt R} \over {\rho ^2}}

& &p^\phi ={1 \over {\rho ^2}}\left( {{a \over \Delta }P-aE+{L \over
 {\sin^2 \theta }}} \right)$$

& &p^\theta ={{\sqrt \Theta } \over {\rho ^2}}
\end{array}
$$

where

$$R \equiv P^2-\left( {m^2r^2+\left( {L-aE}
\right)^2+\wp } \right)\Delta $$

$$P \equiv E\left( {r^2+a^2} \right)-La-eQr$$

$$\Theta \equiv \wp -\left( {a^2\left( {m^2-E^2} \right)+{{L^2}
\over {\sin^2 \theta }}} \right)\cos^2 \theta $$

$$\rho ^2 \equiv r^2+a^2\cos^2 \theta$$

$$\Delta  \equiv r^2-2Mr+a^2+Q^2$$

$$
\begin{array}{llll}
&p^r\equiv m{{dr} \over {d\tau }}
&p^\theta \equiv m{{d\theta } \over {d\tau }}
&p^\phi \equiv m{{d\phi } \over {d\tau }}
\end{array}
$$

and $\wp$ is the Carter constant.

  As before, these expressions were extracted from Misner, Thorne
and Wheeler, 1973 [4]. From these arise the `forces':

$$
\begin{array}{llll}
&F^r={{\partial p^r} \over {\partial r}}{{p^r} \over m}+{{\partial
p^r} \over {\partial \theta }}{{p^\theta } \over m}$$

& &{{dp^\theta } \over {d\tau }}={{\partial p^\theta } \over
{\partial  r}}{{p^r} \over m}+{{\partial p^\theta } \over {\partial
\theta  }}{{p^\theta } \over m}
\end{array}
$$

  Since the results lack the former simplicity and embody no new
physics, it is felt that there is little purpose in listing the
solutions.

\section{Acknowledgements}
    The algebraic facilities made available by the scientific
computing section of ICTP made possible much of the exploration,
although the final results are simple enough to be easily obtained
by hand.

  It is also a tribute to Misner Thorne and Wheeler's `Gravitation'
book that such a calculation can so easily be picked out and
performed.

  Thanks also to Marek Abramowicz, for his patience with the
many varied, and often incorrect attempts at a simplified
derivation of the effect he discovered. That it reduced so
beautifully could not have been possible without the pioneering
work of him and others.

\section{References}

{ \small
1.      M.Abramowicz and J.Lasota, Acta Phys. Pol., {\bf B5}, (1974),
327-.\\ \\
2.      M.Abramowicz and J.Lasota, Acta Astr., {\bf 30}, (1980), 35-.

M.Abramowicz, B.Carter and J.Lasota, Gen. Rel. Grav., {\bf 20},
(1988), 1173-.

M.Abramowicz and A.Prasanna, Mon. Not. R. astr. Soc.,
{\bf 245}, (1990), 720-.

M.Abramowicz and J.Miller, Mon. Not. R. astr. Soc., {\bf 245},
(1990), 729-.

M.Abramowicz, Mon. Not. R. astr. Soc., {\bf 245}, (1990),
733-.\\ \\
3.      S.Chandrasekhar and J.Miller, Mon. Not. R.
astr. Soc., {\bf 167},  (1974), 63-.

M.Anderson, and J.Lemos, Mon. Not. R. astr. Soc., 233,
(1988), 489-.

S.Chakrabarti and A.Prasanna, J. Astrophys. Astr., {\bf 11},
(1990), 29-.

A.Prasanna and S.Chakrabarti, Gen. Rel. Grav., {\bf 22}, (1990),
987-.\\ \\
4.      C.Misner, K.Thorne and J.Wheeler, {\bf `Gravitation'},
Freeman,  1973.
}

\end{document}